# Nuclear Weapons in Regional Contexts: The Cases of Argentina and Brazil[#]

**Olival Freire Junior**

Universidade Federal da Bahia, Brazil – freirejr@ufba.br

**Diego Hurtado**

Universidad Nacional de San Martín, Argentina - dhurtado2003@yahoo.com.ar

**Ildeu C. Moreira**

Universidade Federal do Rio de Janeiro, Brazil - ildeucastro@gmail.com

**Fernando de Souza Barros**

Universidade Federal do Rio de Janeiro, Brazil - fsbarros215@gmail.com

Abstract:

South America is a region which is free from nuclear weapons. However, this was not an inevitable development from the relationships among its countries. Indeed, regional rivalries between Brazil and Argentina, with military implications for both countries, lasted a long time. After WWII these countries took part in the race to obtain nuclear technologies and nuclear ambitions were part of the game. In the mid 1980s, the end of military dictatorships and the successful establishing of democratic institutions put an end to the race. Thus regional and national interests in addition to the establishment of democracies in Latin America have been responsible for the building of trust between the two countries. Meaningful international initiatives are once again needed in the framework of worldwide cooperation. This cooperation is better developed when democratic regimes are in place.

Keywords: Nuclear programs, Nuclear weapons, Brazil, Argentina

---

**Introduction**

South America is a region which is free from nuclear weapons. However, this was not an inevitable development from the relationships among its countries. Indeed, regional rivalries between Brazil and Argentina, with military implications for both countries, lasted a long time. After WWII these countries took part in the race to obtain nuclear technologies and nuclear ambitions were part of the game due to old mistrust and rivalries. However, before these countries managed to acquire such weapons historical events thwarted such ambitions. In the late 1970s, still under military dictatorships, the first signs of an entente emerged with Argentina supporting Brazil on the nuclear treaty with Germany and the agreement among Brazil, Argentina and Paraguay on the use of electricity generated from the waters of shared rivers. In the mid 1980s, the end of military dictatorships, in themselves political phenomena related to Cold War contexts, and the successful establishing of democratic institutions put an end to the race.

Physicists from both countries, through their associations, played a role in this by suggesting and beginning mutual inspections related to nuclear developments. In December 1983 the Brazilian Physics Society (SBF, in Portuguese) and the Argentine Physics Association (AFA, in Spanish), through their presidents Fernando de Souza Barros and Luis Másperi, signed a declaration asking their governments for an agreement concerning collaboration and mutual control on nuclear issues. These initiatives were followed by positions adopted by the new civil governments, under the leadership of José Sarney in Brazil and Raúl Alfonsín in Argentina. Ultimately this led to a treaty between the two countries allowing mutual inspections. Nowadays, restrictions on the use of nuclear technologies for pacific purposes alone figure in the laws of these countries. The case of Brazil and Argentina illustrates the value of scientists' pursuit towards both the pacific uses of nuclear resources and autonomy in nuclear technology, which have been a contentious issue since the end of WWII. This achievement was a result of the historical development of both countries. Its preservation is a challenge for their future histories.

Argentine and Brazilian physicists were not alone in their quest for the peaceful uses of nuclear energy. Indeed they were part of a tradition dating back to the very moment when the first atomic weapons were ready for use in the US, in mid 1945, with Nazi Germany defeated and the war in Europe over. The physicist Niels Bohr approached British and American authorities at the end of WWII to try, unsuccessfully, to convince them to share the new knowledge with the former Allies and the United Nations to foster mutual trust in the new world order (Rhodes, 1986, p. 527) Scientists and technicians at the Metallurgical Laboratory in Chicago led by Leo Szilard petitioned to the US President for wise consideration of the moral responsibilities implied in the use of atomic bombs against Japan.[1] It was also the case of Joseph Rotblat, who left the Manhattan Project at the end of 1944. He objected to its further development once it had become clear that Germany was no longer able to pursue the building of atomic bombs. None of these voices were heeded and the newly created atomic bombs were dropped on the Japanese towns of Hiroshima and Nagasaki as part of the ongoing war in the Pacific. In the second half of the 20$^{th}$ century the world was involved in an arms race, with the US and former USSR leading the dispute, which brought the world to the brink of an unprecedented war. Meanwhile scientists made various attempts to look for control of such weapons. Among such initiatives, the Russell-Einstein Manifesto, the Open Letter to the United Nations, written by Bohr, and the Pugwash Conferences on Science and World Affairs are worth mentioning. Thus, it is not by chance that an early draft of this paper was presented at the international conference "An Open World: Science, Technology and Society in the Light of Niels Bohr's Thoughts", held in Copenhagen 4-6 December 2013.[2] Such a conference seemed to us the appropriate place to bring the experiences of Argentina and Brazil in nuclear issues to the consideration of the international community. The first section of this paper presents a brief review of the history of nuclear physics in both countries. The second section examines the rivalries and the following détente between the two countries. The third section deals with the collaboration and the role played by Argentine and Brazilian physicists in this story while the fourth section presents the agreement of mutual inspections as the main achievement of the two countries. The epilogue reflects on the relationship between nuclear issues and authoritarian regimes and updates the developments in the collaboration between the two countries.

---

[1] Szilard also moved in the same direction as Bohr (Rhodes, 1986, p. 635). The petition led by Szilard is mentioned by Rhodes (1986, p. 697) and its full text is available at http://www.dannen.com/decision/45-07-17.html. Accessed on 10 May 2015.

[2] See http://bohr-conference2013.ku.dk/, accessed on 10 May 2015. On the Open World letter by Bohr in the 1950s, see Aaserud (2007).

**Nuclear science in Argentina and Brazil**

*Argentina*

Physics in Argentina has its roots in the creation of Physics Institute of the National University of La Plata (1906, Buenos Aires province). At this time Richard Gans – former professor of electromagnetism at University of Tübingen – headed the La Plata Physics Institute (1912-1925) from where the first group of Argentine physicists emerged. The leading figure of this group was Enrique Gaviola who had studied in Göttingen and received his doctorate in physics in Berlin in 1926 under the supervision of Peter Pringsheim.[3]

However, the Argentine physics community grew slowly. In the early 1940s, there were around fifteen physicists in Argentina, including advanced graduate students. Interested in attracting top foreign scientists, Gaviola – then director of the National Observatory of Córdoba – was able to obtain permanent residence in Argentina for the anti-fascist Austrian physicist Guido Beck, a former assistant to Heisenberg, who arrived in Argentina in May 1943 (Videira 2001: 160–67). Gaviola and Beck helped set up the Argentine Physics Association (APA) in 1944 and immediately after the news of the atomic explosions in Japan, saw the nuclear issue as an opportunity to promote experimental nuclear physics in Argentina (Hurtado de Mendoza, 2005: 288–91). During the period 1950-51 the Comisión Nacional de Energía Atómica (CNEA) and the Dirección Nacional de Energía Atómica (DNEA) were created. Both institutions focused mainly on the physics of accelerators and reactors, metallurgy, cosmic rays, radiochemistry, and on exploring and prospecting for resources of nuclear raw materials.[4]

At the beginning of 1952, the Argentine government signed a contract with Philips of Eindhoven to purchase a 28 MV synchrocyclotron for deuterons, then a leading instrument for nuclear physics research, and a 1 MeV Cockroft-Walton accelerator, a not-so-modern but easy to use instrument (Mariscotti, 1990: 23). A member of CNEA's laboratory in charge of assembling the machines later wrote:

> [W]e jumped from the typical university laboratory, where an oscilloscope was considered a luxury item, to a shielded steel-and-concrete room with double-walled tanks containing tones of water, and accommodating huge blocks of iron, aluminum, steel, electric generators and control instruments such as we had never seen before in a facility exclusively devoted to scientific research in Argentina (Mayo, 1981: 53).

---

[3] On Gaviola's trajectory, see Bernaola 2001.

[4] After 1956 both institutions were merged.

Crucial organizational initiatives during those early years were the creation of the Metallurgy and Reactor Divisions and the Instituto de Física de Bariloche (today Instituto Balseiro, Río Negro province). Jorge Sabato, a high-school teacher of physics was appointed head of the Metallurgy Division. He was later recognized as the main designer of Argentine nuclear policy and a regional reference on technological policies for development.[5]

*Brazil*

After WWII Brazil had a small but active community doing research in physics. This community had been trained in the 1930s by the Russian-Italian physicist Gleb Wataghin, who had come to São Paulo to set up the physics department at the newly founded Universidade de São Paulo. Their initial work focused on cosmic rays, thus on high energy nuclear physics, and they received international acknowledgment in the late 1930s (Videira and Bustamante, 1993; Freire Jr. and Silva, 2014). The first Brazilian physicists to work on nuclear physics came from this group, among them Cesare Lattes, Marcelo Damy, Paulus Pompeia, and later Oscar Sala. Lattes gained an international reputation as an outstanding experimental physicist after his work on the discovery of pi-mesons. Sala built an electrostatic Van de Graff accelerator in São Paulo with support from the Rockefeller Foundation. Brazilian physicists and chemists were also trained abroad, as was the case of José Leite Lopes, Jayme

---

[5] On Sabato's thinking, see (Sabato 2014).

Tiomno, Hervásio de Carvalho, Sergio Porto, Roberto Salmeron, and Ernst Hamburger, among others.[6]

Government interest in nuclear energy followed the WWII, mainly due to the fact that Brazil had reserves of nuclear fuels such as thorium and uranium. In 1947, during the debates in the Atomic Energy Commission at the United Nations, Brazilian authorities were driven by the statement that,

> Nothing justifies the thesis of a restrictive international policy, capable of summarily depriving nations possessing the raw materials from which nuclear fuels are extracted from the right to utilize them in a peaceful manner, since a similar policy does not apply to other natural sources of hydro energy, also unequally distributed in the several regions of Earth.[7]

Thus there was a confluence of interests among government and scientists which was driven towards research on nuclear science. This common ground was also shared by the military and nationalist politicians and businessmen and helps explain the creation of new research institutes and funding agencies, such as the Centro Brasileiro de Pesquisas Físicas (CBPF), in Rio de Janeiro in 1949, the Conselho Nacional de Pesquisas (CNPQ) in 1951, and the Comissão Nacional de Energia Nuclear (CNEN), in 1956.[8]

**After WWII – Increasing rivalries**

The interest in nuclear physics in Argentina and Brazil was motivated not only by its peaceful uses. The nuclear weapons used by the US in 1945 to force Japan to accept its defeat in WWII provoked worldwide military interest in nuclear technology. Brazil and Argentina joined the race, achieving significant results. Their attempts to acquire the full cycle of nuclear technology were accentuated under military regimes of the 1960-1980 decades. Old mistrust and regional rivalries could become tragedies if nuclear ambitions were to become a part of the old disputes.[9]

---

[6] On Lattes' work, see (Vieira & Videira 2014). On the evolution of Brazilian science in the mid-20$^{th}$ century, see (Schwartzman 1991).

[7] Minutes of the Tenth Session of the Brazilian National Security Council – August 27 1947 – Rio de Janeiro; available at http://digitalarchive.wilsoncenter.org/document/116912. Accessed on 10 May 2015.

[8] On the founding of the CBPF, see Andrade (1998).

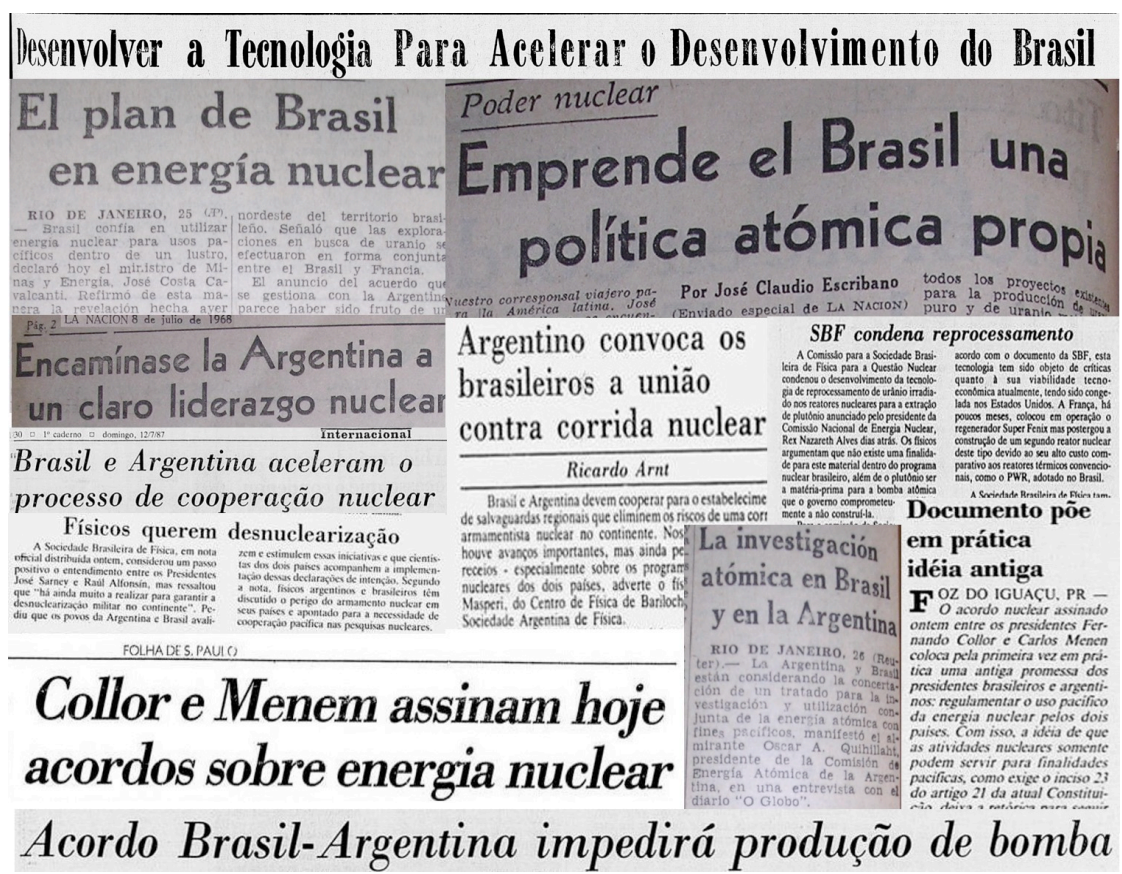

Press coverage reflecting rivalries between Argentina and Brazil

Both Argentina's and Brazil's initiatives were revealed by the military when civilian regimes took over in both countries. During the 1970s, Argentina began a succession of increasingly ambitious projects –none of which finally materialized– at the Ezeiza Research Complex in Buenos Aires. In November 1983, three weeks before Alfonsín took office, the Argentine military government announced the existence of its unsafeguarded gaseous diffusion facility, which had been built by the firm INVAP, a spin-off from Argentina Atomic Energy Commission at Pilcaniyeu (Patagonia). During its first year the plant produced a few tons of slightly enriched uranium. Brazil´s major attempt to create an autonomous nuclear-enrichment facility was a gas centrifuge plant set-up by its Navy at Aramar, with the active collaboration of its Nuclear Energy Commission, and its National Research Council, i.e., with the Brazilian University Research Laboratories in Sao Paulo and Rio de Janeiro. The Aramar facility started operating in 1982, but full disclosure occurred only in 1988, on its public inauguration.[10]

---

[9] Examples of rivalry and mistrust are expressed in the following newspapers: La Nación, 12 August 1967 and 8 July 1968; Jornal do Brasil, 27 November 1990 and 29 November 1990. For a scholarly treatment of the relationship between Brazil and Argentina, see Fausto and Devoto (2004).

**The détente**

While regional competitors, Brazil and Argentina faced the same obstacles any developing country has faced in acquiring nuclear technological autonomy. These circumstances may have led the two countries to a kind of tacit agreement since the late 1960s. Indeed tensions between the two countries enhanced due to another source of energy; hydraulic power. In 1973 Brazil and Paraguay signed a treaty to exploit the Paraná river by building the Itaipu dam in a location not far from Argentina and thus putting the Argentine design of a dam in Corpus at risk. Two years later, Brazil signed a massive agreement with Germany (DDR) to build nuclear plants and for technological transfer, which led to increased American pressure against such an agreement. Argentine diplomacy then moved to support Brazil in international forums.

At that time the Argentine physicist Jorge Sabato claimed that there were good reasons to suspect that the behavior of the US was motivated not for fear of nuclear proliferation but to obtain monopolistic control of the nuclear market. Furthermore, Sabato argued that such international interests had created favorable conditions for collaboration between Brazil and Argentina, which could counter the pressures then on Brazil but in the future on Argentina. Sabato closed his statement citing the

---

[10] On the history of nuclear energy in Argentina and Brazil, see, for instance, Hurtado (2005), Hurtado and Vara (2006), and Andrade (2006) and references herein cited.

Brazilian sociologist Helio Jaguaribe who had written that the key for Latin American independence is agreement between Brazil and Argentina and the key for such an agreement is nuclear cooperation. (Sabato, 1977: 13, 17).

As a matter of fact, the tension about the hydroelectric power plant, Itaipu, on the Paraná river was attenuated and eventually led to the 1979 tripartite treaty among Brazil, Argentina, and Paraguay about the mutual rights on the shared rivers and opened a new era in the relationships between Brazil and Argentina (Redick, 1995: 19-20; Escudé, 1986: 47-49).

**Collaboration among physicists**

In the early 1980s, with the weakening of the military dictatorships in the two countries, physicists from the two countries appealed for collaboration and mutual inspections. In December 1983, the Brazilian Physics Society (SBF, in Portuguese) and the Argentine Physics Association (AFA, in Spanish), through their presidents Fernando de Souza Barros and Luis Másperi, signed a declaration asking their governments for an agreement concerning collaboration and mutual control on nuclear issues (Másperi, 1999: 189). On January 11 of the following year, the AFA published a statement supporting the new President of Argentina Raul Alfonsín and the democratic government to engage itself in the international forums on the pacific use of nuclear energy and interrupt the military interference in nuclear matters. (Regionales, 1984).

Commitments both to the pacific uses of nuclear energy as well as to the challenge of technological autonomy have been long lasting endeavors of Brazilian and Argentine physicists. These date back to the early cooperation between physicists from the two countries in 1945 under the leadership of Guido Beck and Gleb Wataghin. They endured until 1990 when the two countries finally cleared up any remaining doubts about their pacific commitments.[11]

---

[11] Opinião, 03 October 1975; Jornal do Brasil, 23 November 1983, 27 December 1986, 16 July 1987, 29 May 1988 and 19 August 1988. On the cooperation among physicists, (Hurtado y Souza, 2008). For the final pacific commitments, Jornal do Brasil, 27 November 1990 and 29 November 1990.

In July 1984, at the meeting of the Latin American Federation of Physics Societies, held in São Paulo, Brazil, a declaration by the Brazilian, Argentine, and Mexican societies was signed in favor of nuclear disarmament and mutual inspections and controls in Latin America and the Caribbean. Since then, the Argentine and Brazilian societies began to issue a joint annual declaration favoring mutual control on nuclear matters. Such declarations played their role among the public and reinforced the direction that the Brazilian and Argentine governments were working towards a joint nuclear policy (Wrobel & Redick, 2006: 176).

**The agreement of mutual inspections**

Military dictatorships in Latin America ended before the end of the Cold War, which had provided the external context supporting such political regimes. Indeed the days of these dictatorships were counted because of the waning of their legitimacy as consequence of various crises (economic, military, etc) and rising democratic opposition to them. Democracy brought Brazil and Argentina closer and the nuclear issue became a subject of heightened collaboration. In November 30, 1985, presidents Raúl Alfonsin and José Sarney met in Iguaçu Falls to widen collaboration between the two countries. From that meeting the "Joint Declaration on Nuclear Policies" was drawn up and a working team of diplomats, scientists and technicians was created. After 8 months of trading several protocols were signed and Alfonsín invited Sarney to visit the uranium enrichment plant in Pilcaniyeu (Escudé, 1986: 48-49). The following year both presidents visited the Brazilian uranium enrichment plant in Iperó, which resulted from the Brazilian parallel nuclear program under the leadership of the navy and the Argentine plant of reprocessing of plutonium being built in Ezeiza (Ornstein, 1988: 136-140).

     In the late 1980s, Argentine-Brazilian agreements were not weakened with the election of new presidents. In 1990 the Brazilian president Collor de Mello symbolically closed the secretly excavated well supposedly for nuclear military tests in the Amazonian region. In the same year Mello and Menem met each other in Iguaçu Falls renewing the nuclear treaties and the following year representatives from both countries went to Vienna to sign a joint agreement opening their installations for inspection by the IAEA (Goldman, 1991: 9). The far-reaching meaning of these

events was duly noted by contemporary observers. Paul Leventhal and Sharon Tanzer, from the Nuclear Control Institute in Washington, DC, opened the proceedings of a conference dedicated to the theme stating:[12]

> There have been few bright spots in the decades-old, uphill struggle to halt the growth of established nuclear arsenals and to stop the spread of new ones. Nuclear non-proliferation can be a very discouraging business. Thus, it is especially noteworthy when a region troubled by nuclear rivalry – in this case Latin America – beats all the odds and makes a breakthrough toward averting an arms race that most experts regarded as inevitable.

The success of the implementation of these agreements led both countries to sign the Tlatelolco treaty. The history of the Tlatelolco treaty illustrates the tortuous way that led towards a Latin America free from nuclear weapons. Signed in 1967 by countries from Latin America and the Caribbean, it was intended as contribution "towards ending the armaments race, especially in the field of nuclear weapons, and towards strengthening a world at peace."[13] However, for years it risked being a dead letter as Argentina signed it, but did not ratify it, while Brazil signed and ratified it but declared that it only will follow it after all countries had adhered to it. 25 years later, as a consequence of the historical process we are discussing, the treaty was revived, with Argentina and Brazil accepting it in the early 1990s opening the way for Cuba to also adhere to it in the early 2000s. In the meantime, the treaty was updated to preserve the technological achievements of the countries, a requirement introduced by demands from Argentina and Brazil.

---

[12] Leventhal & Tanzer (1992: 1). The conference was held in Montevideo, 11-13 October 1989, and it was titled "Averting a Latin American Nuclear Arms Race – New Prospects and Challenges for Argentine-Brazilian Nuclear Cooperation."

[13] For its full text, see http://www.opanal.org/opanal/Tlatelolco/Tlatelolco-i.htm. Accessed on 12 Nov 2013.

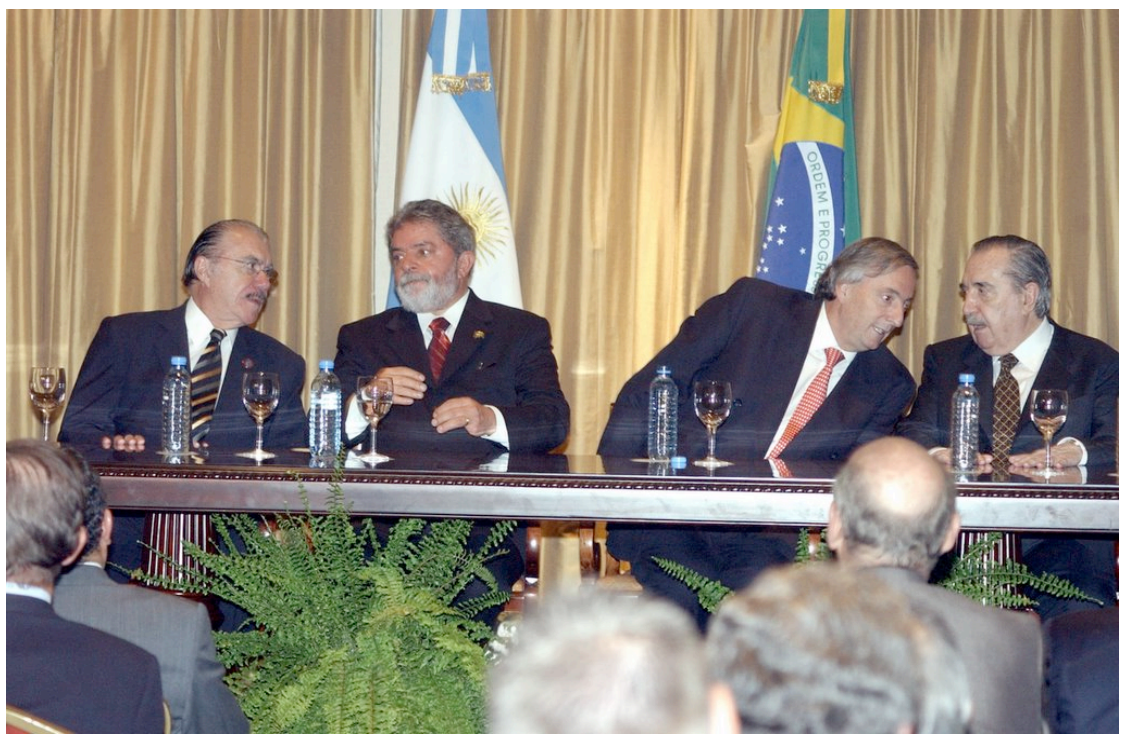

Brazilian and Argentinian presidents, twenty years after the agreement. From the left: Sarney, Lula, Kirchner, and Alfonsin

Democracy was not the sole factor encouraging Brazil and Argentina to commit themselves to the pacific use of nuclear energy. American trade embargoes in the 1980s had motivated Brazilian companies to promote international inspections of Brazil's nuclear facilities. However, the process that led both Brazil and Argentine to accept the UN's international safeguards against nuclear weapons is rather unique. As we have seen the Argentine-Brazil agreement was first established in 1991, when two inspection agencies were set up on the initiative of these countries. Regional easing of tensions was therefore not imposed from abroad. Furthermore, the agreement officially established with the United Nation's inspection agency, the IAEA, established a special set of "inspection rules" to prevent full disclosure of the enrichment facilities. (In the case of Brazil, the acquisition of the blue-print of the centrifuge for the Uranium-235 enrichment and the fast-speed valves were from unrevealed sources.) These "special rules" were accepted by the IAEA because Brazil was exercising its rights within the United Nation's framework.

**Further developments**

The nuclear programs of Argentina and Brazil are no longer a political issue. In both countries nuclear power reactors play a valuable role in electricity supply to the main cities. In the recent years, Brazil and Argentina have agreed to jointly build two nuclear research reactors. The atomic-power agencies from Brazil and Argentina signed an agreement to build two nuclear reactors for research and production of radioisotopes, a Brazilian Multipurpose Research Reactor (RMB) and the RA-10 in Argentina. This agreement meets the Bilateral Integration and Coordination Mechanism, established in the Joint Declaration of 2008 and signed by Presidents Cristina Kirchner and Luiz Ignacio Lula da Silva. The atomic agencies of the two countries have collaborated closely since 2008. Argentina provides Brazil 30% of the Molybdenum 99 (Mo99) radioisotopes which are indispensable in the diagnosis and treatment of cancer.

In both authoritarian and democratic political systems science develops (Freire 2007). According to Heilbron (2003), the general rule is that "science, like most other social activities, does better when encouraged," thus it is insensitive to political systems. The case of nuclear sciences in Argentina and Brazil corroborates this point as these activities were developed in different periods and political systems in these two countries. However, this case also shows us that under democratic systems there may be more social actors interested in the use of nuclear energy for peaceful purposes. This was the case with the democratic governments and scientific societies in both countries.

We may conclude by saying that regional and national interests in addition to the establishment of democracies in Latin America have been responsible for the building of trust between the two countries. Scientists from both countries were influential in obtaining such results. In these countries scientists were committed both to the pacific uses of nuclear energy and to the pursuit of autonomy in nuclear technology, which had been blocked by the countries who first dominated them. The kind of lesson to be drawn from this is related to what lies ahead. The current deterioration of international order could lead to the use of nuclear weapons in politically tense regions. Meaningful international initiatives are once again needed in the framework of worldwide cooperation and this cooperation is better developed when democratic regimes are in place.